\documentclass[a4paper,11pt]{article}
\usepackage{latexsym}
\usepackage{euscript}
\usepackage{epsfig,amsmath,amssymb}

\date{\today}
\def\be{\begin{equation}}
\def\ee{\end{equation}}
\def\bea{\begin{eqnarray}}
\def\eea{\end{eqnarray}}

\long\def\symbolfootnote[#1]#2{\begingroup%
\def\thefootnote{\fnsymbol{footnote}}\footnote[#1]{#2}\endgroup} 

 \large\normalsize

\begin{document}

\begin{center}

{\Large \bf p-q-superstrings in Anti-de-Sitter space-time}
\vspace*{7mm} 

{Betti Hartmann 
\symbolfootnote[1]{{E-mail:b.hartmann@jacobs-university.de}}
and Momchil Minkov}
\symbolfootnote[2]{E-mail:m.minkov@jacobs-university.de  }
\vspace*{.25cm}

{\it School of Engineering and Science, Jacobs University Bremen,
28759 Bremen, Germany }

\vspace*{.3cm}
\end{center}

\begin{abstract}
We study a field theoretical model for p-q-superstrings in a fixed
Anti-de-Sitter background. We find that the presence of the negative cosmological constant
tends to decrease the core radius of the strings. Moreover, the binding energy
decreases with the increase of the absolute value of the cosmological constant.
Studying the effect of the p-q-strings on Anti-de-Sitter space, we observe that the
presence of the negative cosmological constant tends to decrease the deficit angle
as compared to asymptotically flat space-time.
\end{abstract}

\section{Introduction}
The Anti-de-Sitter/Conformal field theory (AdS/CFT) correspondence \cite{adscft} is an 
explicit and well tested realization of the holographic principle which connects
quantum gravity in a $(d+1)$-dimensional space-time to a conformal field theory
on the $d$-dimensional boundary of this space-time. In the case of the AdS/CFT 
correspondence, the $(d+1)$-dimensional space-time is a space-time containing
a negative cosmological constant.
Often it is easier to deal with the lower-dimensional theory on the boundary
than with the full quantum gravity theory.
Recently, the Abelian Higgs model has been studied in a 4-dimensional AdS space-time \cite{mann}. It was found that this model contains string-like solutions. The holographic
description of these solutions via the AdS/CFT correspondence has been studied and it was
found that the space-time around the string contains a deficit angle - very similar
to the case in asymptotically flat space-time.

The interest in string-like solutions has grown again over the past years since it
is believed that cosmic strings might be connected to the fundamental strings
of string theory. The reason is that mechanisms appear in so-called brane world
models \cite{braneworlds} that decrease the fundamental Planck scale, i.e.
the energy scale of fundamental strings, to much lower energy scales.
Moreover, models of brane inflation in which a D-brane and an anti-D-brane collide
and annihilate always seem to predict the creation of networks of cosmic strings
at the end of inflation \cite{braneinflation}.
The objects that form in such brane inflation models are so-called
D-strings and F-strings (for a review see e.g.\cite{pol}). D-strings are 1-dimensional
D-branes and are charged under the Ramond-Ramond potential, while
F-strings are the fundamental strings charged under the Neveu-Schwartz-Neveu-Schwartz
potential. Also  bound states of p F-strings and q D-strings, so-called
p-q-strings are possible. The interesting thing about these bound states is that they are supersymmetric and the square of the tension of these bound states is equal
to the sum of the squares of the tensions of the p strings and the q strings.
Field theoretical models describing these bound states have recently been studied \cite{saffin}.

In this paper, we study a field theoretical model for
p-q-superstrings in  a fixed Anti-de-Sitter background. Our paper is organized
as follows: in Section 2, we give the model, the Ansatz and the equations and
discuss the gravitational effects of p-q-strings. In Section 3, we discuss
our numerical results and we conclude in Section 4.

\section{The model}
The metric of the fixed Anti-de-Sitter background in static, spherical coordinates (representing the coordinates of
an inertial observer)
can be parametrized as follows:
\begin{equation}
ds^2=-\left(1+\frac{r^2}{l^2}\right) dt^2 + \left(1+\frac{r^2}{l^2}\right)^{-1} dr^2  +r^2\left(d\theta^2 + \sin^2\theta d\varphi^2\right)
\end{equation}
where $l=\sqrt{-3/\Lambda}$ is the Anti-de-Sitter radius and $\Lambda$ is the negative cosmological constant.

In the following, we want to study p-q-strings in this fixed Anti-de-Sitter background. The Lagrangian
describing the p-q-strings reads \cite{saffin}:
\begin{equation}
{\cal L}_{m}=-D_{\mu} \phi (D^{\mu} \phi)^*-\frac{1}{4} F_{\mu\nu} F^{\mu\nu}
-D_{\mu} \xi (D^{\mu} \xi)^*-\frac{1}{4} H_{\mu\nu} H^{\mu\nu}
-V(\phi,\xi)
\end{equation} 
with the covariant derivatives $D_\mu\phi=\partial_{\mu}\phi-ie_1 A_{\mu}\phi$,
$D_\mu\xi=\partial_{\mu}\xi-ie_2 B_{\mu}\xi$
and the
field strength tensors $F_{\mu\nu}=\partial_\mu A_\nu-\partial_\nu A_\mu$, 
$H_{\mu\nu}=\partial_\mu B_\nu-\partial_\nu B_\mu$  of the two U(1) gauge potentials $A_{\mu}$, $B_{\mu}$ with coupling constants $e_1$
and $e_2$. The fields 
$\phi$ and $\xi$ are complex scalar fields (Higgs fields).
The potential reads:

\be
V(\phi,\xi)=\frac{\lambda_1}{4}\left(\phi\phi^*-\eta^2_1\right)^2
+\frac{\lambda_2}{4}\left(\xi\xi^*-\eta^2_2\right)^2
-\lambda_3\left(\phi\phi^*-\eta^2_1\right)\left(\xi\xi^*-\eta^2_2\right)
\label{pot}
\ee

In order for $\phi=\eta_1$ and $\xi=\eta_2$ to be the global minimum of this potential, we need to
require that \cite{ahu} 
\begin{equation}
 \lambda_3^2 < \frac{\lambda_1 \lambda_2}{4} 
\end{equation}
When $\phi$ and $\xi$ attain their vacuum expectation values $\eta_1$ and $\eta_2$, respectively, the two $U(1)$ symmetries are spontaneously broken to $1$ and the particle
content of the theory are two massive gauge bosons with masses $M_{W,1}=e_1\eta_1$,
$M_{W,2}=e_2\eta_2$ and two massive scalar fields (Higgs fields) with  masses
$M_{H,1}=\sqrt{2\lambda_1}\eta_1$, $M_{H,2}=\sqrt{2\lambda_2}\eta_2$. 

Each of the two strings consists of a scalar core of radius $r_{H,i}\approx M_{H,i}^{-1}$,
$i=1,2$ and of a magnetic flux tube with radius $r_{W,i}\approx M_{W,i}^{-1}$, $i=1,2$.

\subsection{The Ansatz and equations of motion}
The Ansatz for the matter and gauge fields in spherical coordinates $r,\theta,\varphi$ reads \cite{no}:
\begin{equation}
\phi(r,\theta)=\eta_1 h(r,\theta)e^{i n\varphi} \ \ \ , \ \ \ 
\xi(r,\theta)=\eta_1 f(r,\theta)e^{i m\varphi} \ ,
\end{equation}
\begin{equation}
A_{\mu}dx^{\mu}=\frac {1}{e_1}(n-P(r,\theta)) d\varphi \ \ \ , \ \ \ 
B_{\mu}dx^{\mu}=\frac {1}{e_2}(m-R(r,\theta)) d\varphi \ .
\end{equation}
$n$ and $m$ are integers indexing the vorticity of the two Higgs fields  around the $z-$axis.

Using the above Ans\"atze for the metric and matter fields, the resulting field equations
would be partial differential equations. However, since we want to 
study cylindrical, vortex-type configurations here we can assume that in the following the matter field functions $P$, $f$, $R$ and $h$ depend only on the specific combination $r\sin\theta\equiv \rho$.
The partial differential equations then reduce to ordinary differential equations that depend
only on the coordinate $\rho$ and in the limit $l\rightarrow \infty$ correspond to the equations
of the p-q-strings studied in \cite{saffin}. 

We define the following dimensionless variable:

\begin{equation}
x=e_1\eta_1 \rho \ .
\end{equation}

Then, the total Lagrangian only depends on the following dimensionless coupling constants

\begin{equation}
L=e_1\eta_1 l \ \ , \ \  g=\frac{e_2}{e_1} \ \ , \ 
q=\frac{\eta_2}{\eta_1} \ \ , \ \ 
\beta_i=\frac{\lambda_i}{e_1^2} \ , \ i=1,2,3   \ .
\end{equation}

Varying the action with respect to the matter fields, we
obtain a system of 
four non-linear differential equations. The Euler-Lagrange equations for the matter field functions read:
\begin{equation}
\left(1+ \frac{x^2}{L^2}\right) P'' = 2Ph^2 + \frac{P'}{x}\left(1-\frac{2x^2}{L^2}\right)
\end{equation}
\begin{equation}
\left(1+ \frac{x^2}{L^2}\right) h'' = \frac{\beta_1}{2} h (h^2-1) - \beta_3 h (f^2-q^2) - \frac{h'}{x}\left(1+ \frac{4 x^2}{L^2}\right) + \frac{P^2 h}{x^2}
\label{eqh}
\end{equation}
\begin{equation}
\left(1+ \frac{x^2}{L^2}\right) R'' = 2g^2 Rf^2 + \frac{R'}{x}\left(1-\frac{2x^2}{L^2}\right)
\end{equation}
\begin{equation}
\left(1+ \frac{x^2}{L^2}\right) f'' = \frac{\beta_2}{2} f (f^2-q^2) - \beta_3 f (h^2-1) - \frac{f'}{x}\left(1+ \frac{4 x^2}{L^2}\right) + \frac{R^2 f}{x^2}
\label{eqh}
\end{equation}
where the prime now and in the following denotes the derivative with respect to $x$.

In the general case, the solutions to the above equations have to be constructed
numerically subject to suitable boundary conditions.

The requirement of regularity at the origin leads to the  following boundary 
conditions:
\begin{equation}
h(0)=0, \ f(0)=0 \ , \ P(0)=n \ , \ R(0)=m
\label{eom1}
\end{equation}

In the special case where $n\neq 0$ and $m=0$ 
the boundary conditions (\ref{eom1}) change according to:
\begin{equation}
h(0)=0, \ f'(0)=0 \ , \ P(0)=n \ , \ R(0)=0 \,.
\end{equation}
while for a $n=0$, $m\neq 0$ string, they read
\begin{equation}
h'(0)=0, \ f(0)=0 \ , \ P(0)=0 \ , \ R(0)=m \ .
\end{equation}

By letting the derivatives of the non-winding scalar field 
be zero at the origin instead of imposing the boundary conditions
for the fields themselves, the non-winding scalar field can take non-zero values at the origin and a ``condensate'' of the non-winding scalar field will form in the core of the winding string.

The finiteness of the energy per unit length requires:
\begin{equation}
h(\infty)=1, \ f(\infty)=q \ , \ P(\infty)=0 \ , \ R(\infty)=0  \ .
\end{equation}

\subsection{Asymptotic behaviour}
For $x << 1$, we find that 
\begin{equation}
 P(x<<1)= n + p_0x^2 \ \ , \ \ R(x<<1)=m+r_0x^2
\end{equation}
and 
\begin{equation}
 h(x<<1)=h_0 x^n \ \ , \ \ f(x<<1)=f_0 x^m
\end{equation}
for standard p-q-strings with $n\neq 0$, $m\neq 0$. $p_0$, $r_0$, $h_0$ and $f_0$ are constants that have to be determined numerically.

For a solution with vanishing winding, the behaviour is different. Here, we present the behaviour
for $m=0$ (the $n=0$ case works analogue). We find that $R(x)\equiv 0$
while the corresponding scalar field $f(x)$ forms a condensate at the core of the string
and behaves as
\begin{equation}
f(x<<1)=f_0 + C_1 I_0(\sqrt{A}x) -\frac{B}{A}
\end{equation} 
where $I_0$ is the modified Bessel function of the first kind and 
\begin{equation}
 A= \frac{\beta_2}{2}(3f_0^2 -q^2) + \beta_3 \ \ , \ \
B= \frac{\beta_2}{2}(f_0^2-q^2) + \beta_3 \ .
\end{equation}
$C_1$ is an integration constant. Note that
$f_0=q$ for $\beta_3=0$, while $f_0\neq q $ for $\beta_3\neq 0$.
\\

For $x >>1$, we find a power-law behaviour for the gauge field functions $P$ and $R$ with
\begin{equation}
P(x >> 1) = P_0 x^{c_1} \ \  {\rm with} \ \ 
c_1=-\frac{1}{2}\left(1 + \sqrt{1+8L^2}\right) \ \ , \ \ 
\end{equation}
\begin{equation}
R(x>> 1)= R_0 x^{c_2} \ \ {\rm with} \ \ 
  c_2=-\frac{1}{2}\left(1 + \sqrt{1+8g^2L^2}\right) \ \ .
\end{equation}
where $P_0$ and $R_0$ are constants.

The scalar functions have the following behaviour
\begin{equation}
 h(x>>1)= 1+ H_0x^{\alpha} \ \ , \ \ f(x>>1)= q+ F_0 x^{\alpha}
\end{equation}
where
\begin{equation}
 \frac{F_0}{H_0}=\frac{\beta_1-\beta_2 q +\sqrt{(\beta_1-q\beta_2)^2 + 16q\beta_3^2}}  {4q\beta_3}
\end{equation}
and
\begin{equation}
\label{powers}
 \alpha= -\frac{3}{2} - \frac{1}{2}\sqrt{9+4\beta_1 L^2 - 8 q L^2 \beta_3 \frac{B}{A}} \ .
\end{equation}
For the case where $\beta_1=\beta_2$, $q=1$ (which we will study in this paper) we find - of course - that $H_0=F_0$. Note also that the
positivity of the expression under the square-root is guaranteed by the requirement that
$\beta_3 < \sqrt{\beta_1\beta_2}/2$. 
\\

We define as inertial mass per unit length of the solution the integral
of the energy density $\epsilon=-T_0^0$ over a slice of constant $z=r\sin\theta$.
We have:
\begin{eqnarray}
   \epsilon & = & \eta_1^4\left[  \left(1+\frac{x^2}{L^2}\right)  (h'^2 + f'^2) +
\left(1+\frac{x^2}{L^2}\right)
\left(\frac{P'^2}{2x^2} + \frac{R'^2}{2 g^2 x^2}\right) \right. \nonumber \\ 
      &+& \left. \frac{P^2 h^2}{x^2} + \frac{R^2 f^2}{x^2} + \frac{\beta_1}{4} (h^2-1)^2
+\frac{\beta_2}{4}(f^2-q^2)^2 - \beta_3(h^2-1)(f^2-q^2) \right]  \ .
\label{ed}
\end{eqnarray}

The inertial energy per unit length can then be defined by integrating $T_0^0$
over a section of constant $z$, leading to 
\begin{equation}
             E_{in} = 2\pi \int_0^{\infty} dx \ x \ \epsilon  \ .
\end{equation}
We than define the binding energy of a $(n,m)$-solution as
\begin{equation}
 E^{(n,m)}_{bin}:=E_{in}^{(n,m)}-n E_{in}^{(1,0)}-m E_{in}^{(0,1)}
\end{equation}

\subsection{Gravitational effects}
If we want to study the gravitational effects of p-q-strings on the space-time, we 
have to solve the full Einstein equations. This is very difficult since the resulting
equations would be partial differential equations.
In \cite{mann}, an approximation
for weak gravitational fields and string cores much smaller than 
the Anti-de-Sitter radius was used to study the effects of a single abelian string
on Anti-de-Sitter space-time. In that case, the Einstein equations can be linearized.
We employ this method here for p-q-strings. The metric
reads \cite{mann}:
\begin{equation}
\label{metricads}
 ds^2=\exp(2z/l)\left(-\exp(A)d\hat{t}^2 + d\hat{\rho}^2 + F^2
d\varphi^2\right) + \exp(C)dz^2
\end{equation}
where $A$, $F$ and $C$ are functions of $\hat{\rho}$ and $z$.
For $T_0^0=0$, i.e. in the absence of the p-q-string, the solution to the corresponding Einstein equation would be  $A=C=0$, $F=\hat{\rho}$. The metric (\ref{metricads}) is then just equivalent to the metric
of a pure 4-dimensional Anti-de-Sitter space-time. 

Introducing the rescaled coordinate $\hat{x}=e_1\eta_1\hat{\rho}$ and letting
$z\rightarrow e_1\eta_1 z$, $t\rightarrow e_1 \eta_1 t$, the connection to the
coordinates used before in this paper is given by the following relations \cite{mann}:
\begin{equation}
\hat{x}\exp(z/L)=x\sin\theta \ \ , \ \ \exp(z/L) = \frac{x}{L}\cos\theta + \sqrt{1+\frac{x^2}{L^2}}\cos(t/L) \ \ , \ \ \hat{t}\exp(z/L) = L\sqrt{1+\frac{x^2}{L^2}} \sin(t/L)
\end{equation}
Letting the functions
depend only on the combination $x=\hat{x}\exp(z/L)$, the linearized
Einstein equation for $F$ reads
\begin{equation}
\label{manneq}
 \frac{2}{L^2} - \frac{1}{F} \frac{d}{dx}\left(\left(1+\frac{x^2}{L^2}\right)\frac{dF}{dx}\right)=-\gamma T^0_0
\end{equation}
where $\gamma=8\pi G$ and $T^0_0$ is the energy-momentum tensor of
the p-q-string in the fixed AdS background (see (26)).
The deficit angle $\delta$ of the space-time is then given by $\delta=2\pi(1-F'|_{x=x_0})$, where we assume $T_0^0=0$ for $x > x_0$, i.e.
outside of the core of the string. 
The deficit angle defined in this way thus ``measures'' the departure of the
space-time from a pure Anti-de-Sitter space-time due to the presence of a p-q-string.

\section{Numerical results}
In all calculations, we will fix $g=q=1$ and $\beta_1=\beta_2=2$
and study the properties of the solutions depending on the parameters $\beta_3$, $L$ and on the windings
$n$ and $m$. In all numerical calculations, we have used the ODE solver COLSYS \cite{colsys}.

%%%%%%%%%%%%%%%%%%%%%%%%%%%%%%%%%%%%%%%%%%%%%%%%%%%%%%%%%%
\subsection{Effect of the negative cosmological constant}
%%%%%%%%%%%%%%%%%%%%%%%%%%%%%%%%%%%%%%%%%%%%%%%%%%%%%%%%%%

The presence of a negative cosmological constant tends to decrease the core radii of particle-like objects
as compared to flat space-time. This is also what we observe here. In Fig.\ref{ed}, we plot the energy density
$\epsilon=-T^0_0$
of a $(1,1)$ string for $\beta_3=0.9$ and different values of the parameter $L$ as function of $x/L$.

\begin{figure}[!htbp]
\centering
    \includegraphics[width=11cm]{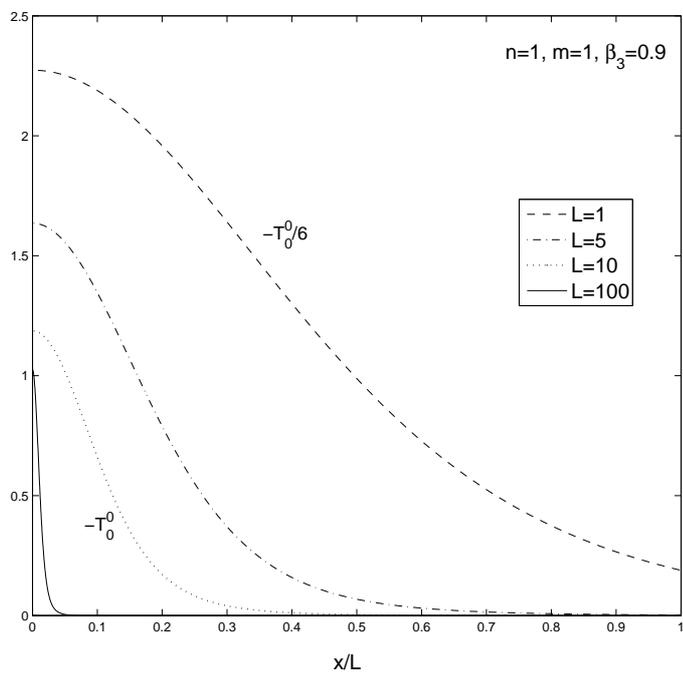} 
\caption{The energy density $\epsilon=-T^0_0$ for $\beta_3=0.9$, $n=m=1$ and
different values of $L$ is shown as function of $x/L$. Note that for $L=1$, we plot $-T^0_0/6$.
 }
\label{ed}
\end{figure}

We observe that for $L$ decreasing, the maximal value of $\epsilon$ increases and at the same
time $\epsilon$ becomes stronger localized \footnote{Note that we plot the energy density as function of $x/L$. If we would plot it as function of $x$ only, the stronger localisation would be apparent.}. The presence of the negative
cosmological constant thus tends to ``squash'' the energy density into a smaller region of space-time,
thus decreasing the core radius of the string. While the effect is small for the Anti-de-Sitter radius $L$ large compared to the radius of the string core, it becomes
much larger when $L$ is comparable to the core radius. Due to our choice of parameters we have (in rescaled coordinates) that $x_{W,i}=x_{H,i}\approx 1$.  For $L=1$, the AdS radius is thus comparable to the 
core radius of the strings.

\begin{figure}[!htbp]
\centering
    \includegraphics[width=11cm]{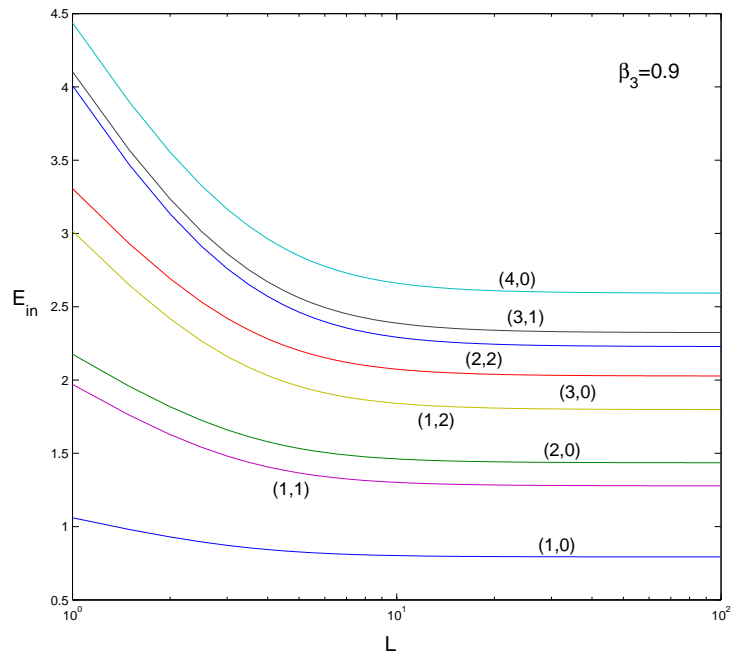} 
\caption{The value of the energy $E_{in}$ is shown as function of $L$ for
$\beta_3=0.9$ and different choices of $(n,m)$.
 }
\label{energy}
\end{figure}

In Fig.\ref{energy}, we show the dependence of the energy (per unit length) of the strings 
on $L$ for $\beta_3=0.9$ and different choices of $(n,m)$.
We observe that the decrease of $L$ leads to an increase of the energy 
and that this increase is stronger for strings with higher total winding $n+m$.
The increase of energy in the case of AdS solitons was already observed in \cite{lms}
for magnetic monopoles.

In addition, we observe that the energy curves run nearly parallel for $n+m$ constant.
The increase in energy due to the negative cosmological constant thus seems - to good
approximation - only to depend on the sum $n+m$ of the two windings.
%%%%%%%%%%%%%%%%%%%%%%%%%%%%%
\subsection{The condensate}
%%%%%%%%%%%%%%%%%%%%%%%%%%%%%

In the particular case where $m=0$, the scalar field function $f(x)$ is not forced to zero at the origin
and can develop a condensate inside the $n$-string. The existence of this condensate was observed before
in \cite{saffin}, but has not been discussed in detail yet. Since it seems that strings
with condensates in their core ``react'' in a particular way to the presence of the cosmological constant,
we study these condensate solutions in more detail here. First of all note that from the potential
(\ref{pot}) we find that a non-zero value of $f$ at $x=0$ can lower the potential energy inside the
string core. The second term in the potential tries to choose a value for
$f$ that is close to $q$, while the third term tries to choose
a value of $f$ close to zero. The potential energy related to these two terms  is minimized (at the origin) for
\begin{equation}
 f(0)=\sqrt{q^2-2\beta_3/\beta_2} \ .
\end{equation}
Of course, this is only an approximation at the origin, while in the full system, we would
have non-linear effects. However, we can already see from this approximation that
$f(0)$ should be decreasing for $\beta_3$ increasing and would be $f(0)=q$ for $\beta_3=0$, i.e.
in the non-interacting case. 
 
\begin{figure}[!htbp]
\centering
    \includegraphics[width=11cm]{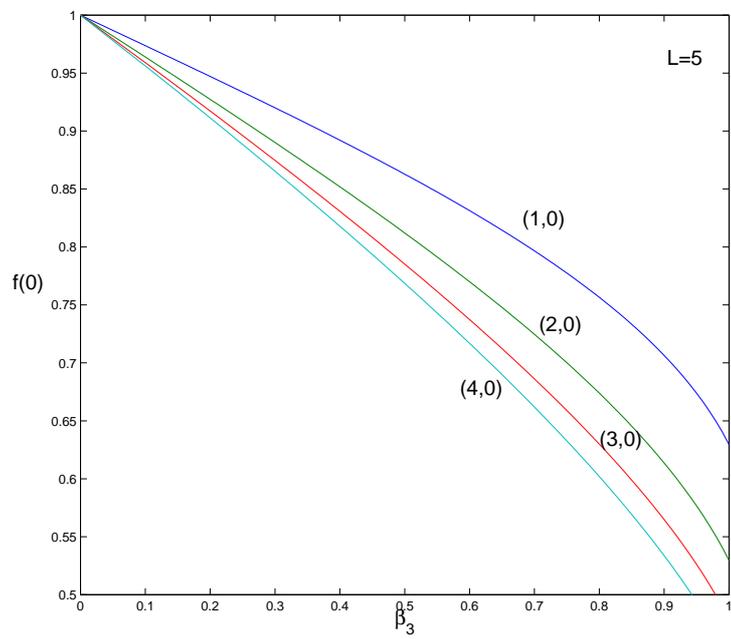} 
\caption{The value of the scalar field function $f(x)$ at the origin, $f(0)$ as function
of $\beta_3$ for $L=5$ and different values of $n$.
 }
\label{f0beta3}
\end{figure}

\begin{figure}[!htbp]
\centering
    \includegraphics[width=11cm]{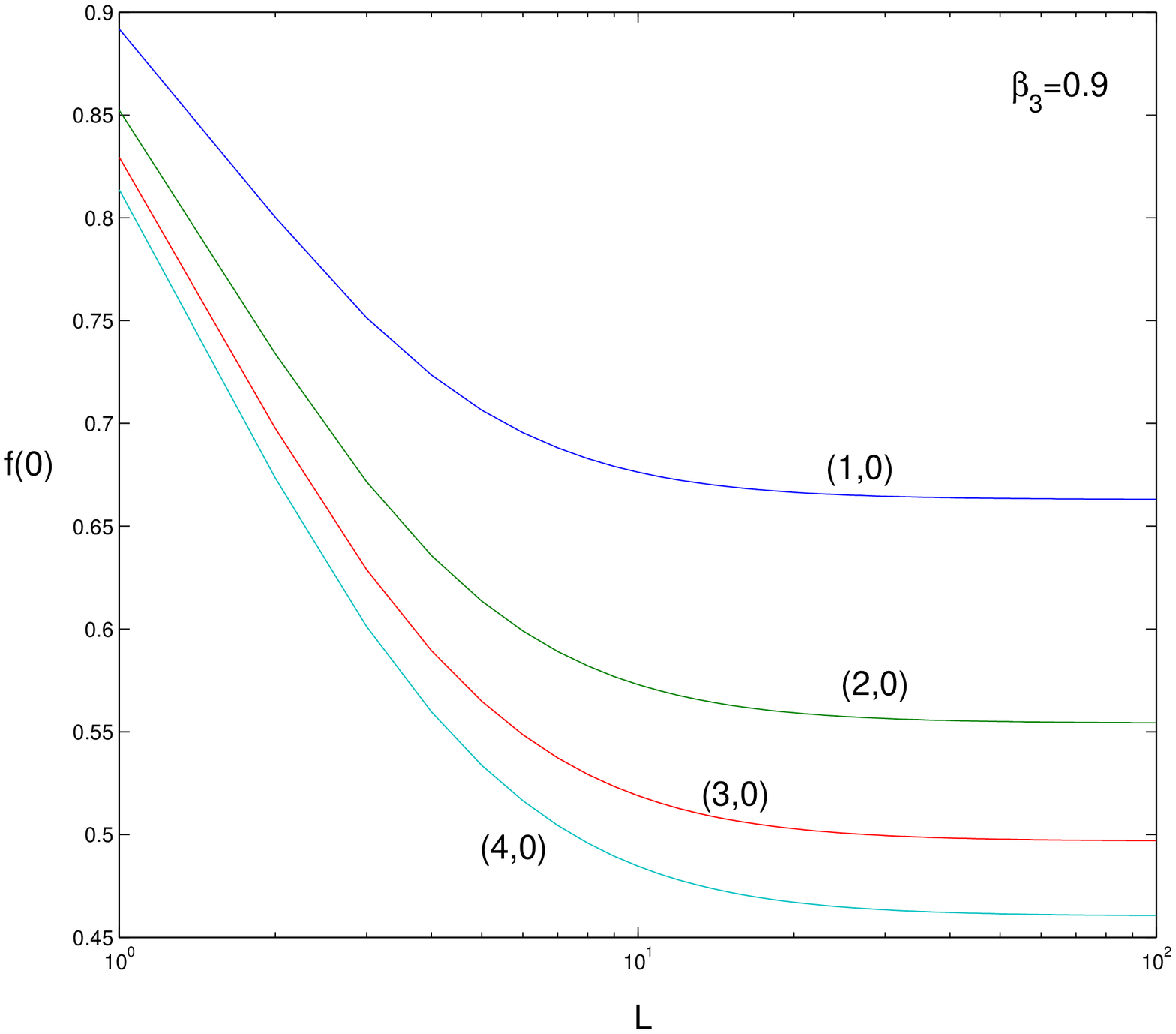} 
\caption{The value of the scalar field function $f(x)$ at the origin, $f(0)$, as function
of $L$ for $\beta_3=0.9$ and different values of $n$.
 }
\label{f0L}
\end{figure}

In Fig.\ref{f0beta3}, we plot the value of the condensate at the origin, $f(0)$, as function of 
$\beta_3$ for $L=5$, that we have determined numerically when solving the system of differential equations. Clearly, the value of $f(0)$ decreases for increasing $\beta_3$ and
tends to $q=1$ in the limit $\beta_3=0$. Moreover, for a fixed value of $\beta_3$ and $L$, the 
condensate decreases for increasing $n$. The reason for this is that for increasing $n$, the core
of the string extends more and more to large $x$. The condensate does the same and has thus
decreasing values of $f(0)$. We also observe that the decrease in the condensate is stronger
going from a $n=1$ solution to a $n=2$ solution than from a $n=2$ to a $n=3$ solution.

In Fig.\ref{f0L}, we show the value of $f(0)$ as function of $L$ for $\beta_3=0.9$.
The curves tend to their non-zero, flat space-time limits for $L\rightarrow \infty$.
Apparently, the value of $f(0)$ increases for decreasing $L$, i.e. increasing absolute value of the 
cosmological constant. The reason for this is the decrease of the core radius mentioned above.
For decreasing $L$ the core radius decreases and the condensate becomes more and more compressed,
thus developing a larger maximal value.

\subsection{Binding energies}

\begin{figure}[!htbp]
\centering
    \includegraphics[width=11cm]{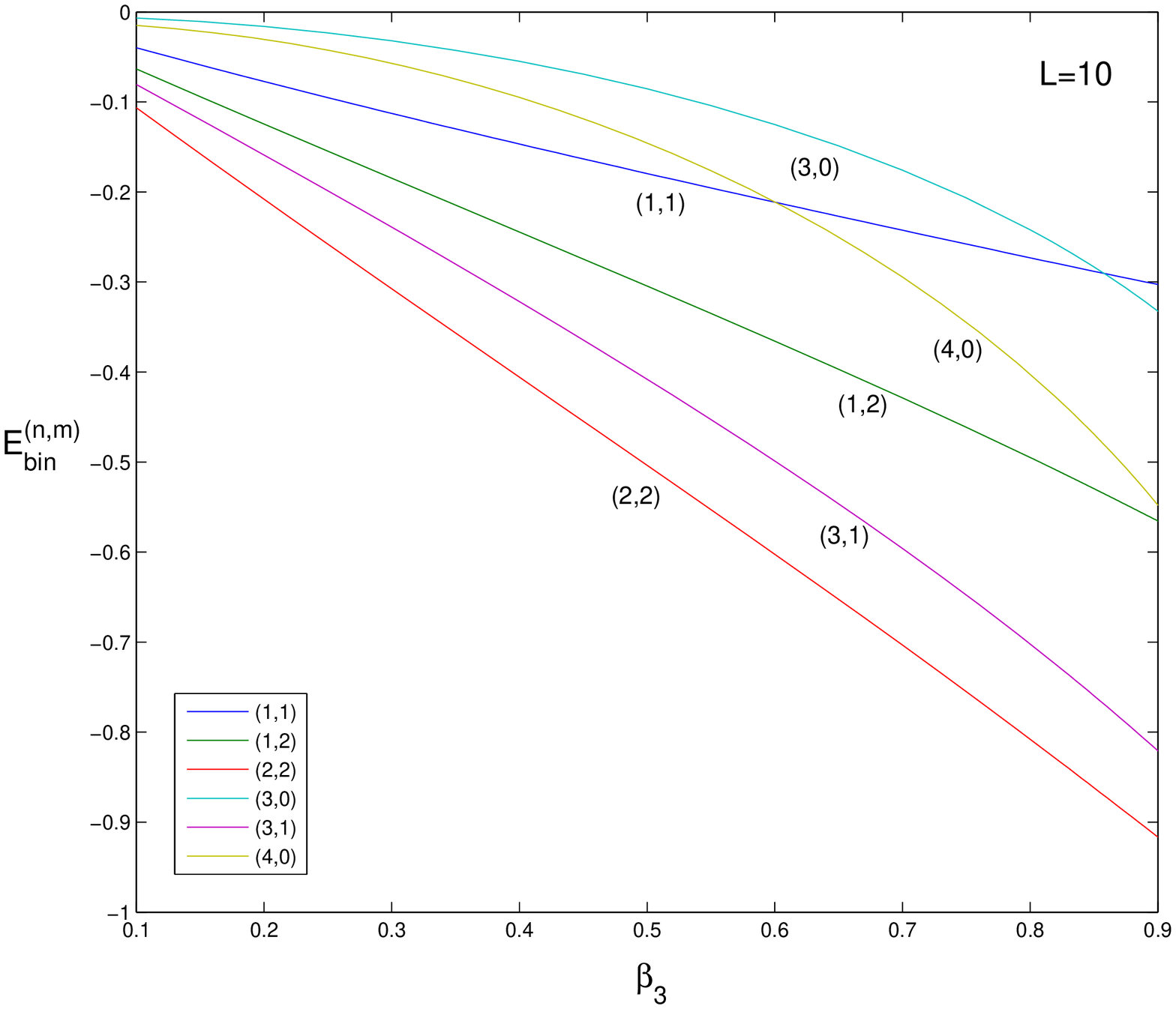} 
\caption{The value of the binding energy $E_{bin}^{(n,m)}$ as function of $\beta_3$ for $L=10$
and different values of $(n,m)$. 
 }
\label{bind1}
\end{figure}

Increasing the parameter $\beta_3$ increases the binding between the strings. We clearly
observe that this is the same in Anti-de-Sitter space. In Fig.\ref{bind1}, we show
the binding energy as function of $\beta_3$ for $L=10$ and different values of $(n,m)$.
The first thing that is apparent from this plot is that the binding energy
curves look qualitatively different for condensate strings with $n\neq 0$, $m=0$
than those of ``standard'' p-q-strings with $n\neq 0$, $m\neq 0$.
While the binding energy seems to decrease approximately linearly with $\beta_3$ for
standard p-q-strings, it decreases to good approximation quadratically with $\beta_3$ for the condensate strings.

\begin{figure}[!htbp]
\centering
    \includegraphics[width=11cm]{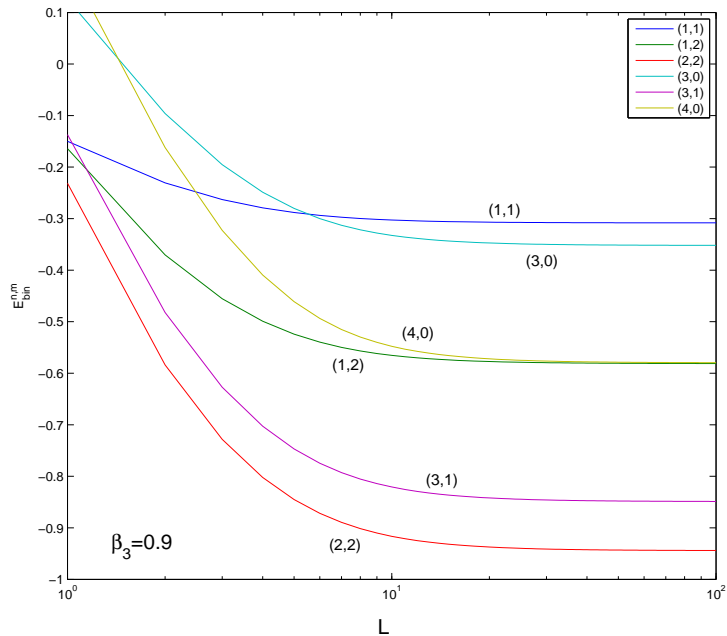} 
\caption{The value of the binding energy $E_{bin}^{(n,m)}$ as function of $L$ for $\beta_3=0.9$
and different values of $(n,m)$. 
 }
\label{bind2}
\end{figure}

In Fig.\ref{bind2}, we show the binding energy as function of $L$ for $\beta_3=0.9$ and different
values of $(n,m)$. As observed in \cite{hu}, the absolute value of the binding energy of a $(n+k,n-k)$ string, $k=0,..,n$
is decreasing for $k$ increasing. We observe the same for Anti-de-Sitter space. E.g. the curves
for the $(2,2)$, $(3,1)$ and $(4,0)$ strings run parallel as $L$ decreases.

The absolute value of the binding energy is decreasing for decreasing $L$, i.e. the larger the absolute value of the 
cosmological constant, the weaker bound are the strings. We even observe that if the 
cosmological constant is large enough, the binding energy becomes positive signalling that the strings become
unbound and would -in a dynamical process - likely separate into $n$ $(1,0)$ and $m$ $(0,1)$ strings. In Table 1, we give $\beta_{3,cr}^{(n,m)}$, the value of $\beta_3$
that separates bound from unbound strings for $L=2$ and different choices
of $(n,m)$.
For $\beta_3 < \beta_{3,cr}^{(n,m)}$, the strings are unbound, while for
larger values of $\beta_3$ they become bound states.

\begin{table}[hbt]
\caption{Values of $\beta_{3,cr}^{(n,m)}$ for $L=2$ and different choices of $(n,m)$}
\label{table1}
\begin{center}
\begin{tabular}{|c|c|c|c|c|c|}
\hline
$\beta_{3,cr}^{(2,0)}$ & $\beta_{3,cr}^{(3,0)}$ & 
$\beta_{3,cr}^{(4,0)}$ &
$\beta_{3,cr}^{(1,2)}$ & $\beta_{3,cr}^{(3,1)}$ &
$\beta_{3,cr}^{(2,2)}$ \\
\hline 
$0.483$ & $0.434$ 
& $0.371$ & $0.033$ & $0.049$ &
$0.040$\\
\hline
\end{tabular}
\end{center}
\end{table}

Clearly, it is the ``condensate strings'' which become unbound for quite
high values of $\beta_3$ and the value of $\beta_{3,cr}^{(n,0)}$ decreases
for increasing $n$. The standard p-q-strings on the other hand become
unbound only for very small values of $\beta_3$. This can be explained
with the different qualitative dependence of the energies of condensate
strings and standard p-q-strings, respectively, on $\beta_3$, which we have discussed before in this paper. It seems that for an AdS radius close to the core radius of the string,
that condensate strings become unbound for quite large values of $\beta_3$.

Moreover, we observe that when $L$ is small enough, a $(1,1)$ string becomes more strongly bound than a $(3,0)$ string
and for even smaller $L$ more strongly bound than a $(4,0)$ string. Thus, standard p-q-strings with a smaller
total winding can become more strongly bound than condensate strings with a larger total winding.  \\

\subsection{Deficit angle}
\begin{figure}[!htbp]
\centering
    \includegraphics[width=11cm]{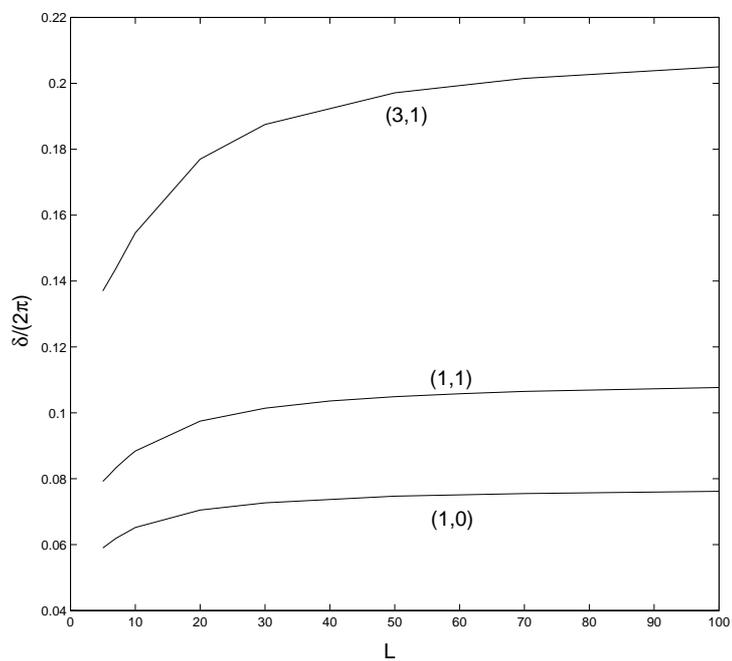} 
\caption{The deficit angle $\delta$ in units of $2\pi$ is given as function
of $L$ for three different choices of $(n,m)$. Here, the gravitational coupling $\gamma=0.1$ and $\beta_3=0.9$.
 }
\label{deficit}
\end{figure}

We have studied the value of the deficit angle by integrating (\ref{manneq}) subject to the boundary conditions $F(0)=0$ and $F'(0)=1$. We have chosen $\gamma=0.1$.
Our results for different values of $(n,m)$ are given in Fig.\ref{deficit}.
For $L\rightarrow \infty$, the value of $\delta$ tends to the value in asymptotically flat space-time. We observe that the deficit angle decreases with the decrease of $L$, i.e.
with the increase of the absolute value of the cosmological constant. Moreover,
the deficit angle decreases stronger for strings with higher total winding $n+m$.
Of course, our calculations are only valid as long as the gravitational field is weak
and as long as the core radius of the p-q-strings is much smaller than the AdS radius, however, we believe that the results of a calculation of the solution
of the ``full'' non-linear Einstein equations would be qualitatively the same.

\section{Conclusions}
We have studied a field theoretical model for p-q-superstrings in a fixed Anti-de-Sitter
background. The negative cosmological constant influences the p-q-strings in different
ways. As observed before for other localized structures, it tends to decrease the core radius of the strings and increases the energy. We observe in particular for condensate
strings which have $n\neq 0$ and $m=0$ that the cosmological constant increases the value of the condensate in the string core. While the model studied here is different from that
describing so-called superconducting strings (in our case, both U(1)s are spontaneously
broken, while in \cite{witten} one U(1) remains unbroken), it would be interesting to
see whether the negative cosmological constant has similar effects on these solutions.

We also observe that if the absolute value of the cosmological constant is large enough
and the interaction parameter is small enough that p-q-strings become unbound.
It would be interesting to know whether this result remains qualitatively the same when studying the
gravitational interaction of p-q-strings with a dynamical AdS space-time and to understand what interpretation this has in the context of the AdS/CFT correspondence.

Studying the gravitational effects within a linearized approximation, we observe that
the space-time has a deficit angle. This deficit angle tends to decrease with increasing absolute value of the cosmological constant. 
\\
\\
{\bf Acknowledgements}
B.H. thanks Y. Brihaye and J. Urrestilla for useful discussions.

\end{document}